\documentclass[a4paper]{article}

\usepackage{INTERSPEECH2020}
\usepackage{multicol}

\title{Asymmetric and trial-dependent modeling:\\the contribution of LIA to SdSV Challenge Task 2}
\name{Pierre-Michel Bousquet, Mickael Rouvier}
\address{LIA - Avignon University}
\email{first.lastname@univ-avignon.fr}

\begin{document}

\maketitle
\begin{abstract}
The SdSv challenge Task 2 provided an opportunity to assess efficiency and robustness of modern text-independent speaker verification systems. But it also made it possible to test new approaches, capable of taking into account the main issues of this challenge (duration, language, ...). This paper describes the contributions of our laboratory to the speaker recognition field. 
These contributions highlight two other challenges in addition to short-duration and language: the mismatch between enrollment and test data and the one between subsets of the evaluation trial dataset.
The proposed approaches experimentally show their relevance and efficiency on the SdSv evaluation, and could be of interest in many real-life applications.
\end{abstract}

\section{Introduction}
The Short-duration Speaker Verification Task 2 evaluation is a text-independent speaker recognition evaluation, based on the recently released DeepMine dataset \cite{deepmine2018odyssey,deepmine2019asru}. This dataset is comprised of various duration utterances (with a significant proportion of less than 10 seconds) recorded by Persian native persons, some of them in English. The evaluation proposes to test and improve speaker recognition methods on speech data with varying degree of phonetic overlap between the enrollment and test utterances~\cite{sdsvc2020plan}. Robustness of speaker embeddings extracted from deep neural networks (DNN) to short-duration utterances and efficiency of the domain adaptation techniques (as Persian language is unknown to the usual speech databases) can be seen as the main objectives of this challenge. The fairly wide DeepMine development dataset provided for this challenge, which is speaker-labeled, allows to better fit model to data, even if the availability of some English speeches spoken by Persian native persons is lacking.

The task of language domain adaptation is usually addressed during the back-end procedure. Several methods have been proposed, unsupervised \cite{Alam18,Bousquet2019,Lee18}, or supervised~\cite{Garcia14ICASSP,Aronowitz2014InterDV,Villalba12} when in-domain labeled data are available. For SdSv, the availability of a relatively large size and labeled in-domain dataset makes it possible to also consider language pre-adaptation inside the supervised learning of a DNN-based feature extractor. Section \ref{section_Frontend} details our proposed approach of DNN Persian-refinement.

\begin{table}[htb]  \centering
  \caption{Data provided by SdSv for speaker verification.}
  \label{tab:TableDataEnrollTest}
  \begin{tabular}{c|c}
  \toprule
    enrollment & test\\
    \hline
    1 to 29 utterances (average of 7) & 1 utterance\\
    giving a net speech duration & (95\% less than\\
    from 3 to 120 seconds & 5 seconds)\\
  \bottomrule
  \end{tabular}
\end{table}
The challenge focuses on short-duration and cross-lingual speaker recognition but it also has a particularity, which is often overlooked in the speaker recognition field: Table \ref{tab:TableDataEnrollTest} shows that the characteristics of the speech material provided for enrollment and for test are different enough to assume a mismatch between the distribution of their vector representations. It would also be of benefit to take into account such a mismatch.
Moreover, mixing, in a unique evaluation, trials with a small or large enrollment sample and, also, test utterances in Persian or English can limit efficiency of a unique modeling. Designing specific back-end models for dealing with trial mismatch could be of interest. Section \ref{section_Backend} explains how we hit on all these points.

\section{Front-end feature extraction}
\subsection{Initial DNN learning}
\label{initial_DNN_learning}
The system used in SdSV Challenge is based on $x$-vector/PLDA. Our $x$-vector system is built based on the Kaldi recipe~\cite{snyder2018x}, but with some modifications. Voxceleb2~\cite{chung2018voxceleb2} and Librispeech~\cite{panayotov2015librispeech} sets are combined to generate the training set for the $x$-vector extractor.

The following data augmentation methods are used in this paper. Apart from the four augmentation methods used in~\cite{snyder2018x}, we also include audio compression randomly picked between ogg, mp3 and flac codec, high-pass filtering randomly picked in [1000Hz;3000Hz] and low-pass filtering randomly picked in [500Hz;1500Hz]. Finally, the training data consist of 8-fold augmentation that combines clean data with 7 copies of augmented data.

During the training part the utterances are further cut into segments of 2s for the neural network training. 60-dimensional filter banks (Fbanks) are used for the $x$-vector system, with an energy-based Voice Activity Detector (VAD) to remove silence. A short-time cepstral mean subtraction is applied over a 3-second sliding window.

Table~\ref{tbl:extended-tdnn} presents the Extended-TDNN architecture used. In addition to this architecture, we proposed to increase the dimension of each layer to 1024 only for the frame-level. Except the layer 9 which is used as an expansion layer and is fixed to 3000 dimension. The embeddings are extracted after the first dense layer with a dimensionality of 512. The neural network is trained for 9 epochs using natural-gradient stochastic gradient descent and minibatch size of 128.

\begin{table}[ht]
\center
\caption{Topology of the Extended-TDNN $x$-vector architecture.}
\begin{tabular}{r c c c}
  \hline
   \textbf{Layer} & \textbf{Layer type} & \textbf{Context} & \textbf{Size} \\
  \hline
  1 & TDNN-ReLU & t-2:t+2 & 1024 \\
  2 & Dense-ReLU & t & 1024 \\
  3 & TDNN-ReLU & t-2, t, t+2 & 1024 \\
  4 & Dense-ReLU & t & 1024 \\
  5 & TDNN-ReLU & t-3, t, t+3 & 1024 \\
  6 & Dense-ReLU & t & 1024 \\
  7 & TDNN-ReLU & t-4, t, t+4 & 1024 \\
  8 & Dense-ReLU & t & 1024 \\
  9 & Dense-ReLU & t & 3000 \\  
  10 & Pooling (mean+stddev) & t & 6000      \\  
  11 & Dense(Embedding)-ReLU & t & 512      \\    
  12 & Dense-ReLU & t & 512      \\
  13 & Dense-Softmax & t & Nb spks      \\        
  \hline
\end{tabular}
\label{tbl:extended-tdnn}
\end{table}

\subsection{Front-end language adaptation}
\label{section_Frontend}
In order to adapt the $x$-vector system to a new language, we use the neural network trained on Voxceleb2 and Librispeech corpus as pre-trained model. Then, we propose to freeze on pre-trained model all pre-pooling TDNN layers and re-train the other layers on DeepMine corpus (using 8-fold augmentation). The neural network is trained only with 1 epoch and minibatch size of 128 (we observe in the leaderboard that more epochs do not improve results).
The resulting "Persian-refined" DNN better combines the rich information of the wide but out-of-domain initial training set and adequacy to the target language.

\section{Back-end asymmetric modeling}
\label{section_Backend}
\subsection{Four-covariance model}
As explained in the introduction and observed in Table \ref{tab:TableDataEnrollTest}, it can be assumed that the distributions of the target speaker model and of the test $x$-vector are sufficiently distinct to require two PLDA modelings. Introduced in \cite{Bousquet2017} for mismatch of duration between enrollment and test data, the four-covariance model (4-cov) is an asymmetric modeling, which allows to compute two distinct PLDA models, here one for enrollment data and the other for test data, then to fit a probabilistic relation between them in order to compute a LLR-score, despite the mismatch.

For SdSv challenge, we choose as target speaker model the length-normalized average of the enrollment sample, since this approach has proved to be efficient and robust~\cite{Rajan2014FromST}. Let denote by $\mathbf{w}_{1}$ a vector of type 1 (here of the latter type, computed on an enrollment sample as described in column 1 of Table \ref{tab:TableDataEnrollTest}) and similarly $\mathbf{w}_{2}$ of type 2 (here a test vector as described in column 2 of Table \ref{tab:TableDataEnrollTest}). The Gaussian PLDA model~\cite{Prince07} for type $i$, $i=1$ or $2$, assumes that:%
\begin{align}
{\mathbf{w}}_{i} &  =\mu_{i}+\mathbf{\Phi}_{i}y_{i}+\varepsilon_{i}\nonumber\\
y_{i} &  \sim\mathcal{N}\left(  0,{\mathbf{I}}\right)  \nonumber\\
\varepsilon_{i} &  \sim\mathcal{N}\left(  0,\mathbf{\Gamma_{i}}\right)
\end{align}

\noindent where $\mathcal{N}$ denotes the normal pdf, $\mathbf{I}$ is the identity matrix, $\mu_{i}$ a global offset and the latent variable $y_{i}$ is only dependent on the speaker and statistically independent of the residual term $\varepsilon_{i}$.
The 4-cov modeling assumes a linear relation between the two PLDA models by their speaker factors:

%\begin{align}
%y_{2}-\mu_{2} &  =\mathbf{A}\left(  y_{1}-\mu_{1}\right)  +\eta\\
%\eta &  \sim\mathcal{N}\left(  0,\mathbf{M}\right)
%\end{align}

\begin{align}
y_{2} &  =\mathbf{A}y_{1}  +\eta\\
\eta &  \sim\mathcal{N}\left(  0,\mathbf{M}\right)
\end{align}

To estimate the matricial parameters $\mathbf{A}$ and $\mathbf{M}$, the point estimate of training speaker factors $y_{i}$ is computed using the expectation given by the PLDA E.M.\ algorithm:%

\begin{align}
y_{i}=\left(  n_{s}\mathbf{\Phi}_{i}^{t}\mathbf{\Gamma}_{i}^{-1}\mathbf{\Phi
}_{i}\mathbf{+I}\right)  ^{-1}\mathbf{\Phi}_{i}^{t}\mathbf{\Gamma}_{i}%
^{-1}\sum_{k=1}^{n_{s}}\left(  w_{k}-\mu_{i}\right)
\end{align}

where $w_{k}$ denotes the $k^{th}$ of $n_{s}$ examples for the speaker $s$.
Then, a multivariate regression is carried out, which minimizes the least square error. Denoting by ${\mathbf{Y}}_{i}$ the row-matrix of the $y_{i}$ the closed-form expressions of $\mathbf{A}$ and $\mathbf{M}$ are:
\begin{align}
{\mathbf{A}}  & ={\mathbf{Y}}_{2}^{t}{\mathbf{Y}}_{1}\left(  {\mathbf{Y}}_{1}^{t}{\mathbf{Y}}_{1}\right)
^{-1}\nonumber\\
\mathbf{M}  & =cov\left(  y_{2}-{\mathbf{A}y_{1}%
}\right)
\end{align}
where $cov()$ is the covariance matrix.
A straightforward computation shows that the LLR score between two vectors $\mathbf{w}_{1},\mathbf{w}_{2}$ of type $1$ and $2$ can be expressed in a simple form (simpler than in the original paper) as:%

\begin{equation}
s\left(  \mathbf{w}_{1},\mathbf{w}_{2}\right)  =-\frac{1}{2}\left(
\begin{array}
[c]{c}%
\mathbf{w}_{1}-\mu_{1}\\
\mathbf{w}_{2}-\mu_{2}%
\end{array}
\right)  ^{t}\mathcal{M}\left(
\begin{array}
[c]{c}%
\mathbf{w}_{1}-\mu_{1}\\
\mathbf{w}_{2}-\mu_{2}%
\end{array}
\right)
\end{equation}
\noindent up to a constant, where

\begin{align}
\mathcal{M} &  =\left(
\begin{array}
[c]{cc}%
\mathbf{\Phi}_{1}\mathbf{\Phi}_{1}^{t}+\mathbf{\Gamma}_{1} & \mathbf{\Phi}%
_{1}\mathbf{\Phi}_{1}^{t}\mathbf{A}^{t}\\
\mathbf{A\Phi}_{1}\mathbf{\Phi}_{1}^{t} & \mathbf{A\Phi}_{1}\mathbf{\Phi}%
_{1}^{t}\mathbf{A}^{t}+\mathbf{\Gamma}_{2}+\mathbf{M}%
\end{array}
\right)  ^{-1}\nonumber\\
&  -\left(
\begin{array}
[c]{cc}%
\mathbf{\Phi}_{1}\mathbf{\Phi}_{1}^{t}+\mathbf{\Gamma}_{1} & \mathbf{0}\\
\mathbf{0} & \mathbf{\Phi}_{2}\mathbf{\Phi}_{2}^{t}+\mathbf{\Gamma}_{2}%
\end{array}
\right)  ^{-1}%
\end{align}

\subsection{Specific score normalization}
\label{specific_S_norm}
Taking benefit of the score normalization to enhance performance requires adapting the usual S-normalization to the specific case of an asymmetric model: the impostor cohorts are dependent on the type of data and the order of pairwise vectors to score must be respected. Given a trial between enrollment-based and test vectors $\mathbf{w}_{e}$ and
$\mathbf{w}_{t}$, score-normalization is performed on score $s\left(\mathbf{w}_{e},\mathbf{w}_{t}\right)$ such that:

\begin{align}
\widehat{s}\left(  \mathbf{w}_{e},\mathbf{w}_{t}\right)    & =\frac{1}{2}%
\frac{s\left(  \mathbf{w}_{e},\mathbf{w}_{t}\right)  -\mu\left(  s\left(
\mathbf{w}_{e},\mathbf{\Omega}_{t}\right)  \right)  }{\sigma\left(  s\left(
\mathbf{w}_{e},\mathbf{\Omega}_{t}\right)  \right)  }\nonumber\\
& +\frac{1}{2}\frac{s\left(  \mathbf{w}_{e},\mathbf{w}_{t}\right)  -\mu\left(
s\left(  \mathbf{\Omega}_{e},\mathbf{w}_{t}\right)  \right)  }{\sigma\left(
s\left(  \mathbf{\Omega}_{e},\mathbf{w}_{t}\right)  \right)  }%
\end{align}
where $\mathbf{\Omega}_{e},\mathbf{\Omega}_{t}$ are cohort impostors, specific to enrollment and test, and $\mu,\sigma$ are the mean and standard deviation functions, possibly computed on the top scores only.

\subsection{Trial-dependent models}
\label{trial_dependent_models}
\begin{table}[htb]
  \caption{Percentages of trials in the evaluation trial set, depending on the target speaker model (how many enrollment segments are available ?) and on the test language.}
  \label{tab:TableFreq}
  \centering
  \begin{tabular}{c|cc|c}
\toprule
& \multicolumn{2}{c|}{test language} &\\
enrollment $\# $segs & Persian & English & Total\\\hline
$< 5$& 36\% & 38\% & 74\%\\
{$\geqslant 5$} & 4\% & 22\% & 26\%\\\hline
& 40\% & 60\% &\\
    \bottomrule
  \end{tabular}
\end{table}
Table~\ref{tab:TableFreq} details the proportion of trials in the evaluation set, depending on the size of the speaker enrollment sample and on the language of test. The 4-cov model allows to fit PLDA models to each of these enrollment-test cases. Table~\ref{tab:TableModels} shows the different training sets used for PLDA, depending on the trial. We apply the 4-cov model to each type of mismatch: (average of sample of various size and duration)/(one short duration utterance in Persian or English).
The language of the test segments is estimated by a speech detector. For test utterances in English, PLDA is interpolated as proposed in~\cite{Garcia14ICASSP}, using our English training database. Let us note that the $x$-vectors of this database are extracted from our Persian-refined neural network, hence partially adapted to Persian language.

For a better understanding, we detail one case of Table~\ref{tab:TableModels}. The last row corresponds to trials with more than 5 examples for enrollment and a test utterance in English:
\begin{itemize}
\item the PLDA training dataset for model 1 of 4-cov model (the one for enrollment) is made up of length-normalized averages of 12 vectors lasting more than 7.5 seconds, extracted from utterances of the DeepMine development set \cite{deepmine2018odyssey}).
\item the PLDA training datasets for model 2 of 4-cov model (the one for test) are comprised of utterances lasting less than 5 seconds, from (i) the same DeepMine development set, (ii) our adapted English development set. The resulting model for test interpolates the last two sub-models (i) and (ii)~\cite{Garcia14ICASSP}.
\end{itemize}

As the final score file to submit mixes four scoring formulas, the scores are calibrated by using development trial datasets, specific to the four cases of Table~\ref{tab:TableModels} and all based on DeepMine development data.

\begin{table}[htb]
  \caption{Datasets for trial-dependent model training. L-average means the length-normalized average of the enrollment sample.}
  \label{tab:TableModels}
  \centering
  \begin{tabular}{cc|c|c}
  \toprule
\multicolumn{2}{c|}{trial:} & \multicolumn{2}{c}{4-covariance model} \\
enrollment & test & model 1& model 2  \\
$\# $segs & language & for enrollment & for test\\\hline\hline

& & 3 vectors & \\
$<$ 5 & Persian & L-average & $<$ 5 sec. \\
& &  $< 5$ sec. &\\\hline

& & 3 vectors & $<$ 5 sec.\\
$<$ 5 & English & L-average & \& \\
& &  $< 5$ sec. & English-dev\\\hline

& & 12 vectors & \\
\multicolumn{1}{c}{$\geqslant 5$} & Persian & L-average & $<$ 5 sec. \\
& & \multicolumn{1}{c|}{ $\geqslant 7.5$sec.} &\\\hline

& & 12 vectors & $<$ 5 sec.\\
\multicolumn{1}{c}{$\geqslant 5$} & English & L-average & \& \\
& & \multicolumn{1}{c|}{ $\geqslant 7.5 $sec.} & English-dev\\

    \bottomrule
  \end{tabular}
\end{table}

\section{Experiments and results}

For acoustic features MFCC are extracted by using Kaldi toolkit~\cite{Povey_ASRU2011} with 23 cepstral coefficients and log-energy, a cepstral mean normalization being applied with a window size of 3 seconds.
Voice Activity Detection removes silence and low energy speech segments. The simple energy-based VAD uses the C0 component of the acoustic feature.
\begin{table}[htb]  \centering
  \caption{Results of the different contributions to the SdSv evaluation.}
  \label{tab:TableResults}
  \begin{tabular}{l|cc}
  \toprule
     & EER\% & minDCF\\
    \hline
    Initial & 7.38 & 0.3682\\
    %\hspace{5pt} + DeepMine dev. & 4.56 & 0.2188\\
    %\hspace{10pt} + refined-DNN & 4.41 & 0.2103\\
    With DeepMine dev. set & 4.41 & 0.2103\\
    \hspace{5pt} + out-of-domain adapted set & 4.42 & 0.1823\\\hline
    \hspace{5pt} 4cov-model& 3.28 & 0.1554\\
    \hspace{10pt}  + specific S-norm& 3.15 & 0.1427\\
    \hspace{15pt}  + trial-dependent models& 2.88 & 0.1261\\
  \bottomrule
  \end{tabular}
\end{table}

Table \ref{tab:TableResults} provides results  of our contributions, in terms of EER and minDCF, as reported in the SdSv Task 2 leaderboard.
The first system (initial) trains the DNN and all the back-end transformations by using only the out-of-domain database, described in section~\ref{initial_DNN_learning}.
The second system benefits from the DeepMine in-domain development set provided by the SdSv organizers. It is used to refine the DNN learning by using the additional training stage described in section~\ref{section_Frontend}, then for learning all the back-end transformations (centering, whitening, length-normalization and PLDA) instead of the initial database.
Let us note that, hence, no adaptation of out-of-domain data to Persian language is carried out during the back-end process to enhance modeling.
The third system additionally leverages an out-of-domain development set for back-end trainings. This dataset is extracted from the one used for the first learning step of the DNN extractor and adapted by using fDA~\cite{Bousquet2019}, an unsupervised domain adaptation method, similar to CORAL~\cite{Alam18}, which takes into account the residual components. The resulting adapted set then allows interpolation between out-of-domain and in-domain PLDA models~\cite{Garcia14ICASSP}.
As expected, systems employing the in-domain development set during front-end and back-end learning outperform the initial submission. It is worth noting that including the adapted out-of-domain development set into the PLDA modelings (row 3) significantly increases performance, but only in terms of minimal DCF.

The fourth system applies the four-covariance model. Let us note that this system does not use the adapted out-of-domain dataset during the back-end trainings. For the enrollment model, the training speaker models are the length-normalized averages of 15 examples (3 original segments + 12 data augmented) and, for the test model, only training segments of less than 5 seconds are selected.
The gain of performance involved by this method is significant, both in terms of EER and minDCF, even without the help of the wide out-of-domain development set.

The following system adds to the latter the specific score-normalization proposed in section \ref{specific_S_norm}, with 400 top-scores.
The resulting gain of performance shows that the normalization of score is compatible with an asymmetric model.

The last system applies the trial-dependent 4-cov modeling described in section \ref{trial_dependent_models} and Table \ref{tab:TableModels}.
The gain of performance confirms the heterogeneity between the trial partitions listed in Table \ref{tab:TableModels} and the ability of the 4-cov model to handle such type of mismatch.

Relevance and efficiency of our various contributions are clearly demonstrated. The final system takes full account of the challenges of SdsV Task 2: short-duration utterances and adaptation to new language, reported in terms of performance in the first rows of Table \ref{tab:TableResults}, then mismatch between enrollment and test distributions or trial partitions, reported in the last rows.

\section{Conclusions}
The SdSv challenge made it possible to test and compare the efficiency of DNN based systems to deal with short-duration utterances. Data augmentation could also contribute to better fit these data, which are known to be very varied.
The task of language adaptation was usually tackled during the back-end process. For the SdSv challenge, the availability of a sizable in-domain labeled dataset allowed to extend this task to the DNN supervised learning stage.

Our contribution highlights the concern of mismatch between enrollment and test speech material, in terms of quantity of information. The proposed four-covariance model applies a specific asymmetric modeling, which focuses on a type of mismatch. It reveals the benefit of refining the back-end modeling to take into account this issue. Moreover, this model allows for better fit of specificities, here the relative heterogeneity of the evaluation trials.

The last system of Table \ref{tab:TableResults} was our final submission for this challenge.
The good ranking obtained with a system using a single front-end feature extractor shows that a system including all these contributions is able to compete with fusions of systems based on distinct DNN architectures and configurations.
\section{Acknowledgements}

This research was supported by the ANR agency (Agence Nationale de la Recherche), on the RoboVox project (ANR-18-CE33-0014).

\bibliographystyle{IEEEtran}

\bibliography{mybib}

\end{document}